\documentstyle[prl,aps,epsfig,floats]{revtex}

\newcommand\beq{\begin{equation}}
\newcommand\eeq{\end{equation}}
\newcommand\bea{\begin{eqnarray}}
\newcommand\eea{\end{eqnarray}}

\newcommand\nonum{\nonumber}
\newcommand\sqpi{\sqrt{\pi}}
\newcommand\tphi{{\tilde\phi}}

\newcommand\on{\omega_n}
\newcommand\bi{\begin{itemize}}
\newcommand\ei{\end{itemize}}

\begin{document}

\draft

\textheight=23.8cm
\twocolumn[\hsize\textwidth\columnwidth\hsize\csname@twocolumnfalse\endcsname

\title{\Large Transport through multiply connected quantum wires}
\author{\bf Sourin Das and Sumathi Rao } 
\address{\it   
Harish-Chandra Research Institute,
Chhatnag Road, Jhusi, Allahabad 211019, India}
\date{\today}
\maketitle

\begin{abstract}

We study transport through multiply coupled carbon
nano-tubes (quantum wires) and
compute the conductances through the two wires as a
function of the two gate voltages $g_1$ and $g_2$ controlling the
chemical potential of the electrons in the two wires.
We find that there is an {\it equilibrium}
cross-conductance, and we obtain its dependence on the 
temperature and length of the wires.
The effective action of the model for the wires in the strong  coupling 
(equivalently Coulomb interaction) limit can also be 
mapped to a system of capacitively coupled quantum dots.
We thus also obtain the conductances  for identical and
non-identical dots.
These results can be experimentally tested.

\end{abstract}

\vskip .5 true cm

\pacs{~~ PACS number: 71.10.Pm, 73.23.Hk, 73.63.Kv}
\vskip.5pc
]
\vskip .5 true cm

\section{Introduction}

Transport in one-dimensional systems (quantum wires) has
continued to attract interest in the last decade. This 
has been  mainly
due to the fabrication of 
novel one-dimensional materials like single-walled carbon nano-tubes,
besides the more standard quantum wires obtained by gating semi-conductors.
Moreover, attention has been attracted by the evidence for  Luttinger
liquid behavior in the non-linear transport measurements
on these carbon nano-tubes\cite{bockrath,yao,bae}. 
This has led to an upsurge of theoretical work\cite{kbf,eg,many} 
on transport through carbon nano-tubes.

Transport measurements involving more than one carbon nanotube
can show even more dramatic deviations from Fermi liquid behaviour.
For instance, the predictions\cite{ke}  for crossed carbon nano-tubes 
have been experimentally verified\cite{kim}. Further predictions\cite{ke2}
have been made for longer contacts leading to Coulomb drag as well.

In this paper, we study a system of two carbon nano-tubes with
a slightly different geometry. The aim is to understand the
phenomena of resonant tunneling through coupled carbon nanotubes.  
We start with a system of two wires with 
density-density couplings operating at the two ends of both the 
wires\cite{pdn}.
This geometry is also relevant in the study of entangled electrons,
where a superconductor (source of entangled electrons)
is weakly coupled to the wires and the consequent nonlocal correlation 
is measured at the two edges\cite{samuelson}. 
The experimental situation that we wish to analyze is given in Fig.(1).
The carbon nano-tube wires are between the source and the
drain and the two floating gates above the wires provide
a strong capacitive coupling between the two wires at
both ends. We will see later that  
the same setup can also be thought of as a set of two quantum dots in 
parallel, at least in the strong interaction  limit, where the 
density-density couplings between the wires themselves are the tunnel 
barriers responsible for forming the dot. ( We will see this
analogy explicitly in the effective action.) 

\begin{figure}[t]
\begin{center}
\epsfig{figure=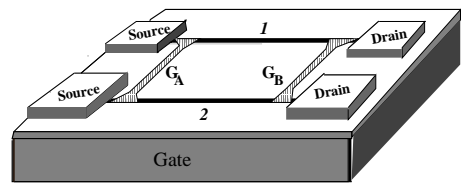,width=6.0cm}
\end{center}
\caption{Carbon nanotube wires 1 and 2 are stretched
between the source and the drain. $G_A$ and $G_B$  are  the two floating gates 
which generate a strong capacitive coupling between the two  
wires at the two ends.}
\end{figure}

Our aim is to  compute the conductance through the two wires 
as a function of the two gate voltages $g_1$ and $g_2$
controlling the density  of electrons in wires $1$ and $2$. 
Without leads, when the
electron-electron Coulomb interaction strength is weak,
the capacitive coupling between the wires gets renormalized 
to zero and  the system decouples
into two independent wires. The resonant transmission pattern
in this case is well-known and is simply the resonant transmission
between double barriers. But in the limit of strong inter-electron 
interactions, the coupling between the wires grows, 
and an interesting resonance pattern
emerges, With the inclusion of leads, we find that the
value of the interaction strength for which the coupling changes
from being irrelevant to relevant changes. With leads, even
stronger inter-electron interactions are needed to access the
strong coupling regime. Fortunately, for carbon nano-tubes, 
the interaction parameter is in this regime, and 
the interesting resonance pattern
that  emerges in the strong coupling limit  can be experimentally tested.

In Sec. (II), we show how  the system can be modelled 
in terms of the one-dimensional
bosonised Luttinger liquid Lagrangian. In Sec. (III), we  obtain the effective
action by integrating out all degrees of freedom, except at the
coupling points, firstly for a uniform wire 
without leads.  We then show that for identical wires, the
system decouples in terms of symmetric (`+') and antisymmetric ($`-'$)
combinations of the fields at the coupling points (boundary fields),
and we are essentially left with two copies of a wire with  
back-scattering potentials at the two ends.
In the  weak Coulomb
interaction limit, the back-scattering couplings potentials
renormalise to zero (at very low temperatures $T\rightarrow 0$ and
long wire lengths $d \rightarrow \infty$). 
However, for strong Coulomb interactions, the back-scattering potentials
turn out to be  relevant.
At $T\rightarrow 0$ and  $d\rightarrow \infty$, the wires are `cut'
and there is no transmission. However, there is still the possiblity
of resonant transmission. When mapped back to the original wires,
the conductance maxima forms an interesting resonance pattern.
For non-identical wires, there is a coupling 
term, which can
be treated perturbatively and the change in the pattern of 
resonance maxima can be  explicitly obtained.
In both the cases of strong and weak Coulomb
interactions,  the conductances through the two wires can be 
explicitly computed in terms of the new $`+'$ and $`-'$ fields  
perturbatively. In the `high' temperature limit, the temperature
$T$ is the scale of the cutoff of the renormalisation group (RG)
equations and the conductances are a function of $T$. In the low temperature 
limit, and for finite length wires, the length of the wire is the
RG cut-off, and the length dependences of the conductances 
can be obtained. Interestingly, 
we find that there is a non-zero {\it equilibrium cross-conductance}
- $i.e.$, there is a non-zero current through one wire caused by
a voltage drop across the other wire.
Finally, we show that the inclusion of leads changes the value
of the interaction parameter where the coupling between the wires
changes from being irrelevant to relevant.
The resonance patterns
do not change due to the inclusion of leads, but the dependence
of conductances on the temperature and the length of the wires,
which depend on the RG flow of the coupling
strengths do change and we compute the new conductances.

In Sec. (IV), we show how the effective action in the 
strong interaction limit is identical to the action that one
would get for two capacitively coupled quantum dots. Hence, we show
that our results are also applicable to a system of capacitively
coupled quantum dots. Finally in Sec. (V), we conclude with 
a discussion of  how the
current model can be extended to multiply coupled wires and
multiply coupled dots.

\begin{figure}[htb]
\begin{center}
\epsfig{figure=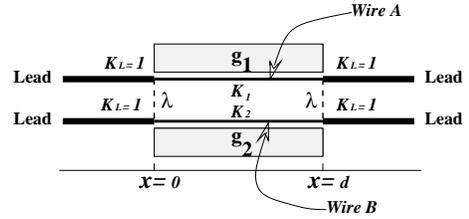,width=6.0cm}
\end{center}
\caption{Schematic diagram of the experimental setup. The wires 
are modeled as LLs with $K_{i =1,2}$, with the leads having
$K_L=1$. The density-density coupling between the wires is 
denoted by its strength  $\lambda$ and the two gate voltages
controlling the densities of electrons in the wire are
denoted by $g_1$ and $g_2$.}
\end{figure}

\section{The Model}

Following Ref.\cite{kbf}, we will assume that the band structure
of the carbon nanotube is captured by the one-dimensional 
free fermion model given by
\beq
H_0 = -\sum_{i,\alpha}~\int dx ~v_F~  
[\psi_{Ri\alpha}^{\dagger}i\partial_x \psi_{Ri\alpha} - 
(R\leftrightarrow L)] 
\eeq
where $i=1{,}2$ refers to the two wires and $\alpha=\uparrow,\downarrow$ 
refers to the two spins. 
Coulomb repulsion between the electrons can be approximated
as an onsite density-density interaction as follows -
\beq
H_{int} = \sum_{i} ~\int dx ~ \rho_{i\uparrow} \rho_{i\downarrow}~. 
\eeq
Here  
$\rho_{i\alpha}(x) = \psi_{i\alpha}^{\dagger}\psi_{i\alpha}$
are the electron densities of the $\uparrow$ and $\downarrow$
electrons  and $\psi_{i\alpha} = \psi_{Li\alpha}~
 e^{-ik_F x} +\psi_{Ri\alpha}~e^{ik_Fx}$. The $\psi_{Ri\alpha}$ 
and $\psi_{Li\alpha}$ 
stand for fermion fields linearized
about the left and right Fermi points in the $ i^{th}$  wire .
Using the standard bosonisation procedure,
whereby a fermionic theory can be rewritten as a bosonic theory
with the identification  $\psi_{i\alpha} =(\eta_{i\alpha}/\sqrt{2\pi
\alpha})
~e^{2i\sqpi \,\phi_{i\alpha}} $, 
the Hamiltonian  can be written as
\beq
H = H_0 +H_{int}  =\sum_{i\beta}
\frac {v_{i\beta}}{2}\bigg[K_{i\beta}(\Pi_{i\beta})^2 
-\frac{1}{K_{i\beta}}(\partial_x\phi_{i\beta})^2\bigg]
\eeq
where $\beta = c,\sigma$ are the subscripts for charge and spin 
degrees of freedom, instead of $\alpha = \uparrow,\downarrow$ since
the interaction term mixes them. $\eta_{i\alpha}$ are the Klein
factors that ensure the anti-commutation relations of the fermions.
Here $K_{ic} \sim(1+g/\pi v_F)^{-1/2}$ , $v_{ic}\sim v_F(1+g/\pi v_F)^{1/2}
$ and~ $\Pi_{i,c/\sigma}$~ are  the fields dual to~ $\phi_{i,c/\sigma}$.
$K_{i\sigma} =1$ and $v_{i\sigma} = v_F$ in the absence of magnetic fields.
$K_{ic}=1$ for  free electrons  and $K_{ic} < 1$
for repulsive   $e$-$e$   interactions.
For simplicity, we will now drop spin indices and work with  
spinless  electrons,
and write down the action for the setup in Fig. 1 as 
\beq
S = \int d\tau~ [~L_{{leads}} + L_{{wires}} + L_{ {coup}}+ 
L_{ {gates}}~]~.  
\label{fullaction}
\eeq
The electrons in the leads are free while the electrons in wires 1 and 2 are 
interacting and they are modelled as Luttinger liquids with 
Luttinger parameter $K_{iL}=1$ and  $K_i=K_1, K_2$ respectively - 
\bea
L_{{leads}} +L_{wires}&=& \sum_{i=1}^2
[(L_{{leads}})_i ~+~ (L_{wires})_i] \nonumber \\
&=& \sum_{i=1}^2\bigg(\int_{-\infty}^{0} + \int_{d}^\infty \bigg) 
dx {\cal L}_i(\phi_i;~K_{iL},~v_{F})   \nonum\\
&+& \sum_{i=1}^2
\int_{0}^{d} dx~ 
 {\cal L}(\phi_i;K_i,v_i).
\eea

Here $\phi_i$ denotes the (spinless) Luttinger
bosons in wires 1  and 2 respectively
with the Lagrangian densities  
\beq
{\cal L}_i (\phi_i;{K_i},{v_i})
 = (1/2 {K_i})[ (1/v_i) {(\partial_t\phi_i)}^2 - 
(v_i){(\partial_x \phi_i)}^2 ]~.
\eeq
The Lagrangian for the coupling 
between the wires is given as
\bea
L_{ {coup}} &=&   
\int_{- \infty}^{+ \infty} dx~ [~
\lambda_1 ~\rho_1(x)~\rho_2(x) ~\delta(x) ~+\nonum \\
&& ~~~~~~~~~~~~~~ 
\lambda_2~\rho_1(x)~\rho_2(x) ~\delta(x-d)~]\nonum\\
&=&
\frac{\lambda}{(\pi\alpha)^2}~
\bigg[ \cos({2\sqrt\pi \phi_1^1})
\cdot \cos({2\sqrt\pi \phi_2^1})
\nonum \\
&+&
\cos({2\sqrt\pi \phi_1^2 + 2 k_F d})
\cdot 
\cos({2\sqrt\pi \phi_2^2 + 2 k_F d}) \bigg] ~,
\eea
where $\rho_i$ are  the densities of the electrons, and in terms of the 
bosonic fields, they are  given by 
\bea
\rho_i(x)= {1\over \sqpi} \partial_x \phi_i(x) + {1\over \pi \alpha}
\cos(2\sqpi \phi_i(x) + 2 k_F x). 
\eea
Here $\alpha$ is an infra-red regulator 
and we have set Klein factors to 1 ( which is sufficient for the
correlation functions we compute in this paper, although 
in general with two wires, one has to be careful.)
The gate voltage that couples to the electrons densities in the two
dots is modeled by the following term in the action -
\beq
L_{ {gates}} = \sum_{i=1}^2 g_i \int_{0}^{d} dx~ \rho_i(x) =
\sum_{i=1}^2 {g_i \over \sqrt \pi} (\phi_i^2 - \phi_i^1) ~. 
\eeq

\section{The effective action}

\noindent $\bullet$ {\bf Uniform wire with no leads}
\vspace{0.5cm}

We first analyse the model of a uniform quantum wire with
\beq
S = \int d\tau ( L_{wires} + L_{coup} + L_{gates}),
\eeq
where $L_{wires} = \sum_{i=1}^2 \int_{-\infty}^\infty 
{\cal L}_i (\phi_i; K_i,v_i)$.
The terms in the Lagrangian for the coupling between the wires 
$L_{coup}$ and the coupling between the wires and the gates,
$L_{gate}$, are both  functionals only 
of the fields at the boundaries ($i.e$. at $x=0$ and $x=d$).
Hence, even in $L_{wires}$ 
(which is a  functional of the bulk fields), 
it is convenient for further calculations to integrate out all 
degrees of freedom except at the coupling points, $x=0$ and $x=d$,
and obtain an effective action\cite{kf} 
in terms of the boundary fields  given by
\bea
(S_{ {eff}})_0 &=& \int d\tau\sum_{i=1}^2(L_{wires})_i
\nonum \\ 
&=& \sum_{i=1}^2 \frac{1}{2K_i} 
\int {\vert \omega \vert}
( {({\tphi_i^1})}^2 + {({\tphi_i^2})}^2) d\omega \nonum \\
&+& \sum_{i=1}^2 
\frac{1}{2K_i} 
\int \frac{\vert \omega \vert}  { e^{k_i d} -
e^{-k_i d}}
~ \Big\{ (e^{k_i d} + e^{-k_i d}) \nonum \\ \nonum \\
&& 
( {({\tphi_i^1})}^2 + {({\tphi_i^2})}^2)  
 - 4 {\tphi_i^1}{\tphi_i^2}
\Big\} \, d\omega~. 
\label{Seff0}
\eea
The Fourier transformed tilde fields are defined by 
$
\phi_i^{1,2}(\tau) = \sum_{\on} e^{-i\on \tau} \tphi_{i}^{1,2} (\on).$
In the high frequency or equivalently, high temperature limit, 
($T \gg \hbar v_i /k_B d$),
$(S_{eff})_0$ reduces to 
\beq
(S_{eff})_0^{ht} = \sum_{i=1}^2 
\frac{1}{K_i} 
\int {\vert \omega \vert}( {({\tphi_i^1})}^2 + {({\tphi_i^2})}^2)
d\omega~.
\label{ht}
\eeq
In the low temperature  limit  
($T \ll \hbar v_i /k_B d$), 
the density-density coupling at the two ends of 
each wire are seen coherently by the electrons
and the total effective action reduces to 
\bea
S_{ {eff}} 
&=& (S_{eff})_0^{lt}+\int d\tau ~[L_{coup} + L_{gates}]
\nonum \\ &=& \sum_{i=1}^2 
\frac{1}{2K_i} 
\int {\vert \omega \vert}( {({\tphi_i^1})}^2 + {({\tphi_i^2})}^2)~
d\omega \nonum \\
&+&
\int d\tau \bigg[ \sum_{i=1}^2 {U_i \over 2} 
(\phi_i^2 - \phi_i^1 )^2 
\nonum \\ 
&+&  
L_{coup} +L_{gates} ]~. 
\eea 
The $U_i = \hbar v_i/K_{i} d$ are the mass terms
that suppress charge fluctuations on the wires  and are responsible
for the Coulomb blockade (CB) through the wires. 
In this paper we  consider the symmetric case where 
$\lambda_1=\lambda_2=\lambda$ (say). 

For further analysis, it is convenient to define the following 
variables -
\bea
&\theta_{i}& = {\phi_i^1 + \phi_i^2 \over 2} + {k_F d\over 2},~~  
{N}_{i} = {\phi_i^2 -\phi_i^1 \over \sqpi} + {k_F d\over\pi}~. 
\eea
In terms of these variables the coupling part of the action can be written as 
\bea
{L}_{{coup}}&=& {\lambda \over \pi \alpha}
[\cos 2\sqpi(\theta_{1} +\theta_{2}) 
\cos 2\sqpi ({N}_{1}+{N}_{2}) \nonum \\
&+& 
\cos 2\sqpi(\theta_{1} -\theta_{2}) \cos 2\sqpi 
({N}_{1}-{N}_{2})]~.
\eea
The expression for $L_{coup}$ in terms of the new fields suggests that 
we can diagonalise the problem by introducing the following 
symmetric and anti-symmetric combination of fields -
\beq
\theta_{\pm} = \theta_{1} \pm \theta_{2}, 
~~ {N}_{\pm} = {N}_{1} \pm {N}_{2}~.
\eeq
Thus, the total action (in the low $T$ limit) can be written 
in terms of these new $\pm$  fields as
\bea
S_{ {eff}} 
&=& \int d\tau ~ \bigg[\sum_{\nu =+,-} [
\frac{{\cal U}_{eff}}{2} \, ({ N}_{\nu} - { N}_{0\nu})^2]
\nonum \\
&+&~
\frac{{\cal U}_1-{\cal U}_2}{4}\, { N}_{+} \,{ N}_{-} \nonum \\
&+& 
{\lambda\over (\pi \alpha)^2}~ \sum_{\nu=\pm}
[\cos2\sqpi\theta_\nu \cos\pi { N}_\nu]\bigg]~,
\label{ea}
\eea
where ${\cal U}_{eff} = ({\cal U}_1 + {\cal U}_2)/4,
{\cal U}_i =\pi \,{\rm U}_i$ and 
${S_l}$,~${N}_{0+}$ and ${N}_{0-}$ are given by
\bea
{S_l}~
&=&
\frac{1}{2}~{\sum_{\on}}\bigg[(\tilde { N}_+)^2
+\frac{\pi}{4}{(\tilde \theta_+)}^2 
+ (\tilde { N}_-)^2
+\frac{\pi}{4}{(\tilde \theta_-)}^2 \bigg] 
\label{newlead} \\ &&{\rm and}~~ \nonum \\
{ N}_{0\pm} 
&=& 
\frac{{ k_F d}\,({ {\cal U}}_1\ \pm  { {\cal U}}_2) - 
({g_1}\pm{g_2})} {(\pi/2) ({{\cal U}}_1 + {\cal U}_2)}~.  
\eea
Thus,  from Eq.(\ref{ea}), we see that when ${\cal U}_1
={\cal U}_2$, we have successfully mapped the 
problem of two density coupled quantum wires to a 
pair of ``decoupled'' quantum wires  with double barriers.
When ${\cal U}_1 \ne {\cal U}_2$  but is  {\em small}, 
the two wires interact weakly.
It is also possible to identify effective 
Luttinger liquid parameters for the $`\pm'$ wires from ${\cal U}_{eff}$
(for ${\cal U}_1 = {\cal U}_2$) by writing it as ${\cal U}_{eff}= \pi U_{eff} =
\pi \hbar v/K_{eff} d$  and we find that 
$K_{eff} = 4K_1 K_2/(K_1+K_2)$ for both the `+' and
`$-$' wires, since both of them have the same ${\cal U}_{eff}$.

Note that for $K_1=K_2=K$, $K_{eff} = 2K$. Hence,
the interaction parameter has doubled\cite{ke}.
The $`+'$ and $`-'$ wires are `free' when $K=1/2$, and the
quasi-particles have repulsive  interactions for $K<1/2$ and attractive 
interactions for $K>1/2$.  
\\

\noindent {\bf {(a) Case of ${\cal U}_1={\cal U}_2={\cal U}$ :}}  \\

When $U_1=U_2$  the coupling term between the '$+$' and '$-$'   fields  
drops from the effective action and the action is  exactly identical 
to the effective 
action of a  decoupled pair of quantum wires, each with 
two barriers, in one dimension\cite{kf}. 
The action remains invariant under the following transformations 
\beq
\theta_\pm \rightarrow \theta_\pm + \frac{\sqpi}{2}, 
~~N_\pm \rightarrow 2N_{0\pm} - 
N_\pm~,
\eeq
\noindent 
when $N_{0\pm}$ is tuned to be a half-integer by  tuning the 
gate voltages $g_1$ and ${g_2}$. 
This  tuning of gate voltages corresponds to certain special 
points in the  (${g_1},{g_2}$) plane where resonance transport of electrons
through each of   
the wires  takes place. 

The conductance matrix for the two-wire system. in the linear 
response regime, can be
written as
\beq
G = \left( \begin{array}{ll}
   G_{11} ~& G_{12} \\
   G_{21} ~& G_{22} \\
 \end{array} \right)~.
\eeq
where $G_{ii}$ is the conductance through each wire due to the
voltage across the same wire and $G_{ij}$ is the cross-conductance - 
the conductance in wire $i$ due to the voltage drop in wire $j$.
{\it Note that the density-density couplings at the two ends of the
wires can be thought to be the source of `entanglement' of
the previously uncorrelated electrons in wires 1 and 2.} The 
cross-conductance $G_{12}$ is a measure of this entanglement.

By transforming to the $`+'$ and $`-'$ wires, we can compute 
$G_{\pm}$  explicitly since they are just the
conductances for uncoupled wires with two barriers each. Moreover, since
$G_{\pm}$ can be written in terms of
the currents  $j_\pm = j_1 \pm
j_2$, we find that
\bea
G_+ (g_1+g_2)&=& \bigg(G_{11} +G_{22} +G_{12}+G_{21}\bigg)
(g_1,g_2) \nonumber \\  G_- (g_1-g_2)&=& \bigg(G_{11}+G_{22} -G_{12}-
G_{21}\bigg)(g_1,g_2), 
\eea
from which we can obtain 
\bea
G_1+G_2 &=& G_+ + G_-  = G \\
G_{12} + G_{21} &=& G_+ - G_-~.
\label{cond}
\eea
Note that $G_{12} +G_{21} \ne 0$, as long as $G_+ \ne G_-$.
Thus unlike in the case of the singly crossed carbon 
nano-tubes\cite{ke}, where the cross-conductance $G_{12}$ was
a non-equilibrium effect and vanished in the linear response regime,
here, {\it the cross-conductance is an equilibrium phenomena}.
It only requires $G_+ \ne G_-$. Since the conductances $G_+$
and $G_-$ are independently tuned by the gate voltages
$g_1 +g_2$ and $g_1-g_2$, they are equal only if they are
both tuned to be equal.

The combined transport through the wires 1 and 2 can now be
tuned to resonance when both $G_+$ and $G_-$ are tuned
to resonance (maximum),  or when one of them is tuned
to resonance (semi-maximum). It is easy to see that the  
conductance maxima through both the  dots form a 
rectangular grid of points in the plane of the gate voltages,  
when  $N_{0+}$ and $N_{0-}$ are tuned to 
half integers - $i.e.$, $N_{0+} = n+1/2$ and $N_{0-} = m+1/2$.
The values of the appropriate gate voltages, given below,  
\bea
g_2 + g_1  &=& \big[\frac {2 k_F d}{\pi} - 
(n+{1\over 2})\big]~{\cal U} \nonumber \\
g_2 - g_1  &=& \big[(m+{1\over 2})\big]~{\cal U}
\eea
are plotted in Fig. 3. The semi-maxima form the two pairs of solid lines
and the intersections of the two pairs of lines are the maxima.
    
\begin{figure}[htb]
\begin{center}
\epsfig{figure=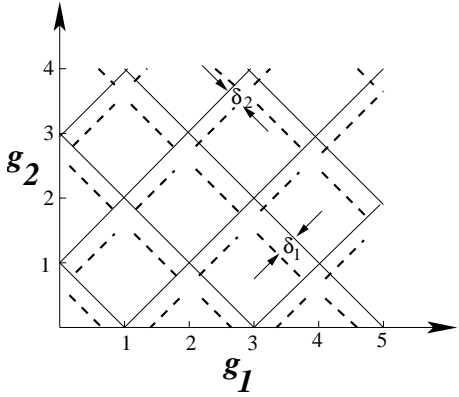,width=6.0cm}
\end{center}
\caption{Conductances in the plane of the two gate voltages $g_i$.
For ${\cal U}_1 = {\cal U}_2$, the solid lines represent semi-maxima 
($`+'$ or $`-'$  wire at resonance) and 
the crossings represent the maxima (both wires at resonance). The dotted 
line represents 
semi-maxima ($`+'$ or $`-'$  wire at resonance) for ${\cal U}_1 \ne{\cal U}_2$.
There is no resonance at the crossings of the dotted lines
and that region has been left blank.}
\end{figure}
\vspace{0.25cm}

\noindent{\em{(i) Weak interaction ,  weak coupling limit :}}  
\\

After transforming to the 
$`\pm'$  wires, the RG flow of the  
density-density coupling term is given by 
\beq
{d\lambda\over d l} =  (1-K_{eff})\lambda  .
\eeq
For weak inter-electron interactions $(K >1/2$ or $K_{eff} >1)$,
the  the  $\lambda$ coupling is irrelevant and it  
grows smaller as a function of the energy
cutoff $l = {\rm ln}[\Lambda(\lambda)/\Lambda]$. Here,
$\Lambda$ is an arbitrary high energy scale (say, the inverse of the
average inter-particle separation)
at which we start the renormaisation group flow.
We are interested in the conductance of the system in the
low temperature limit ($T \ll T_d$) where 
there is coherent transport
through both the barriers. The total conductance through 
the system is  given by\cite{nagaosa}
\bea
G &=&\frac{2e^2}{h} - \frac{\alpha}{2} e^2 \lambda^2  
(\frac{T}{\Lambda})^{2(K_{eff}-1)}\times \nonum \\ 
&& [2+\cos2{\pi}N_{0+}+\cos2{\pi}N_{0-}], 
\eea
where, $\alpha$ is an arbitrary constant of order unity.
Note that the last
factor ($i.e.$, factor in square bracket)
goes to zero when both the wires are tuned to 
resonance and then there is perfect conductance through both
the wires.

Similarly, the temperature dependence
of the total cross-conductance is  given 
by 
\bea
G_{12} +G_{21} &=& \frac{\alpha}{2} e^2 \lambda^2 
(\frac{T}{\Lambda})^{2(K_{eff}-1)}
\nonum \\ && (\cos2{\pi}N_{0+}-\cos2{\pi}N_{0-}) 
\eea 
in the low temperature limit. Thus, the
cross-conductance is non-zero unless both wires are at
resonance or $N_{0+} = N_{0-} + 2\pi N$ and it can be both 
negative or positive depending on the gate voltages ($g_1$,$g_2$) operating 
on the two wires.
Note also that as the temperature is reduced, the effective barrier
strength reduces and hence, perturbation theory
is a good approximation. In fact, for $T\rightarrow 0$, 
the direct conductances are very close to 
$e^2/h$ or perfect conductance and the 
cross-conductances go to zero. 
\\

\noindent{\em{(ii) Strong interaction ,  strong coupling limit :}}  
\\

For strong inter-electron interactions $(K_{eff}<1)$ or $K<1/2$, the 
density-density coupling term is relevant under the RG
transformation. 
At low energies, 
the strength of the barriers  $\lambda$  renormalize 
to very large values, and in fact, at zero temperature, or for very 
long wires, the $`+'$
and $`-'$ wires are cut and there is no transmission, except
at resonance points. However, for 
finite temperatures and for finite length wires, there are 
power law conductances given by\cite{kf}
\beq
G \sim e^2 t^2 
(\frac{T}{\Lambda})^{2(1/K_{eff}-1)} \sim G_{12} +G_{21}
\eeq
where $t$, of order $1/\lambda$, is  a tunneling amplitude 
between the cut wires.
In this limit, the system can be considered as a pair of decoupled
dots, ($`+'$ and $`-'$ dots) which are tunnel-coupled to Luttinger wires
(see Fig. 6). The density-density couplings which grow themselves act
as the barriers forming the dot.

\vspace{0.5cm}
\noindent $\bullet$ {\bf Wire with leads}
\vspace{0.5cm}

Let us now incorporate the leads, by studying the model in 
Eq.(\ref{fullaction}). The leads have  non-interacting electrons
and only the electrons within the length of the
wire between the two coupling terms are interacting.

The inclusion of leads essentially changes the renormalisation
group flow of the barriers in the two wires\cite{nagaosa,safi,sid}. 
To find the new
RG flows, we first note that even with the inclusion of leads,
it is convenient to work with the $`+'$ and $`-'$ fields, where
the two wires decouple. In terms of these fields, we find
that the interaction parameters in the leads are also changed
and are given by $K_{\pm L} = 2$ as compared to $K_i=1$.
Hence, the leads are no longer `free'. 

When the barrier
is at the boundary between the leads and the wire as in our
set-up in Figs 1 and 2, the RG equations are given by\cite{sid}
\begin{displaymath}
{d\lambda\over d l} = \left\{ \begin{array}{ll} 
                 (1-K'_{eff})\lambda  & T \gg T_d \\ \\
                 (1-K_{\pm L}) & T \ll T_d 
                    \end{array} \right.
\end{displaymath}
where $K'_{eff} =2~ K_{eff} K_{\pm L}~/~(~K_{eff} +K_{\pm L}~)
= 4K/(K +1)$ and $T_d= \hbar v/k_B d $ as before. So for the 
'$\pm$'  wires, $\lambda$ is relevant if   
$K'_{eff} < 1$,~$i.e$.~ $K  <  1/3$ and 
it is irrelevant  for  $K'_{eff}  >  1$,~$i.e$.~ $K  >   1/3$. \\

\noindent{\em{(i) Weak interaction , weak coupling limit :}}   \\ 
For  `weak'  inter-electron interactions $(K'_{eff}>1)$ or $K>1/3$, the 
density-density coupling term is irrelevant under the RG
transformation. Note that connecting  leads to the 
interacting  wires, changes the 
values  of  $K$  for which the density-density coupling is  irrelevant  
from   $K \geq 1/2$  to  $K \geq 1/3$~. So we  observe that  even if  
$K<1/2$, but $K>1/3$, the  density-density coupling still remains  
irrelevant,  unlike the  case of  uniform wire with no leads.

The high 
temperature conductance scales now with $K'_{eff}$ instead of 
$K_{eff}$. But 
at low temperatures ($T \ll T_d$) where there is coherent transport
through both the barriers, 
there exists a  new feature. The conductance now has
both non-trivial temperature and length dependences and is  given by
\bea
G &=&\frac{2e^2}{h} - \frac{\alpha}{2} e^2 \lambda^2 ({T\over T_d })^{
2(K_{\pm L} -1)} 
(\frac{T_d}{\Lambda})^{2(K'_{eff}-1)}\times \nonum \\ 
&& (2+\cos2{\pi}N_{0+}+\cos2{\pi}N_{0-}) 
\eea
The temperature dependence essentially comes because in the
$`\pm'$ wires, the leads are no longer free.

Similarly, the temperature and length dependences
of the total cross-conductance is given 
by 
\bea
G_{12} +G_{21} &=& \frac{\alpha}{2} e^2 \lambda^2 ({T\over T_d })^{
2(K_{\pm L} -1)} 
(\frac{T_d}{\Lambda})^{2(K'_{eff}-1)}
\nonum \\ && (\cos2{\pi}N_{0+}-\cos2{\pi}N_{0-}) 
\eea
This result is experimentally testable and is plotted in Fig.(4).

\begin{figure}[htb]
\vspace{0.2cm}
\begin{center}
\epsfig{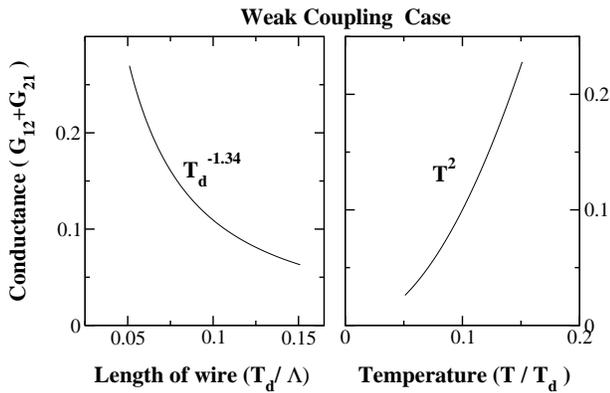}
\end{center}
\caption{ The dependence of the low temperature ($T \ll T_d$) 
cross-conductance (in units of $2e^2/h$)
on $T$ and $T_d$ for the case of $K=.5$.~Here $T_d=\hbar v/k_B d$  
is the temperature equivalent of the length $d$ of the wire. 
$\Lambda$ is the high energy  cutoff scale. The overall scale of
the conductance has been adjusted by adjusting  $\alpha$, to be 
within the perturbative regime. Hence, it is only the power law
which is significant.}
\end{figure}
\vspace{0.25cm}

\noindent{\em{(ii) Strong  interaction , strong coupling limit :}}    \\

For  strong   inter-electron interactions $(K'_{eff}<1)$ or $K<1/3$, the 
density-density coupling term is relevant under the RG
transformation. Connecting  leads to the interacting  wires changes 
the regime of 
 $K$  for which the density-density coupling is  relevant  
from   $K < 1/2$  to  $K < 1/3$~.\,  
So from an experimental point of view,
carbon-nanotubes are very good candidates for testing our predictions 
in this limit, as its  Luttinger  parameter($K$) ranges from 
0.2 to 0.3\cite{bockrath,yao,bae}. 
Since $\lambda$  renormalises  to very large values in this regime, for very 
long wires or at very low  temperature, the $`+'$ and $`-'$ wires are cut and 
there is no transmission, except at resonance points. 
However, for finite temperatures and for finite length wires, 
as usual, we can compute the conductances as a function of the
temperature and/or length scale perturbatively.
The high  temperature ($T \gg T_d$) 
conductance scales now with $1/K'_{eff}$, instead of $K'_{eff}$ as in
the weak interaction case.  At low temperatures ($T  \ll  T_d$)
also, the interaction parameters get replaces by their inverses\cite{kf,sid}
and  the direct and cross-conductances are  given by
\beq
G \sim e^2 t^2 ({T\over T_d })^{2 (1/K_{\pm L} ~-1)} 
(\frac{T_d}{\Lambda})^{2(1/K'_{eff}~-1)} \sim G_{12} +G_{21}
\eeq
where  $'t'$  is the  tunneling  between the cut wires.  
This result is also experimentally testable and is plotted in Fig.(5). \\  

\begin{figure}[htb]
\vspace{0.2cm}
\begin{center}
\epsfig{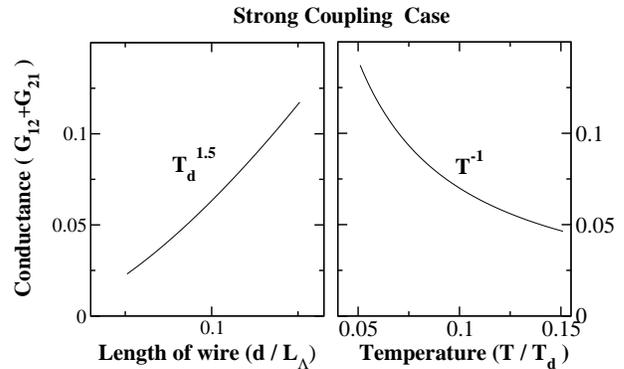}
\end{center}
\caption{ The dependences of the low temperature ($T\ll T_d$) 
conductance (in units of $2e^2/h$) 
on $T$ and $T_d$ for the case of $K=.25$.}
\end{figure}
\vspace{0.25cm}

\noindent{\bf (b) Case of ${\cal U}_1\ne {\cal U}_2$ :}  \\

However, when ${\cal U}_1 \ne {\cal U}_2$, 
it is no longer possible to tune for resonances through the 
$`+'$ and $`-'$ wires
simultaneously due to the presence of the $N_{+}N_{-}$  term in 
the effective action. 
$ N_{0+} $  now depends 
on  $ N_- $ and $ N_{0-} $ depends on $ N_+ $.
But it is always possible to fix either $N_{0-}$ or $N_{0+}$ and 
tune the other wire to resonance. The condition
for resonance for the $`+/-'$  wire is given by
 \beq
V_{eff}(N_{\pm},\theta_{\pm};N_{\mp},\theta_{\mp} ) =  
V_{eff} (N_{\pm}+1,\theta_{\pm} +\frac {\sqrt{\pi}}{2};
 N_{\mp},\theta_{\mp} )
\eeq
and the appropriate gate voltages at which  the wires
  get tuned to resonance is given by
\bea
g_1 + g_2  &=& \bigg[\frac {2 k_F d}{\pi} - \big(n+\frac{1}{2}\big) 
-\frac{\pi}{4}~\bigg(\frac{{\cal U}_1 - {\cal U}_2}{{\cal U}_1+{\cal U}_2}
\bigg) ~   \bigg]~{\cal U}_{eff} 
\nonumber \\
g_1 - g_2  &=& \bigg[\frac {2 k_F d}{\pi}\bigg(\frac{{\cal U}_1-{\cal U}_2}
{{\cal U}_1+{\cal U}_2}\bigg) 
- \big(m+\frac{1}{2}\big)
\nonumber\\
&-&
\frac{\pi}{4}~\bigg(\frac{{\cal U}_1 -  
{\cal U}_2}{{\cal U}_1+ {\cal U}_2} ~ \bigg)  \bigg]~{\cal U}_{eff}, 
\eea
which gives us the deviations $\delta_1$ and $\delta_2$ in Figs 3
and 4 as
\bea
\delta_1 &=& \frac{\pi}{4}~\bigg(\frac{{\cal U}_1 -  
{\cal U}_2}{{\cal U}_1+ {\cal U}_2} ~ \bigg) {\cal U}_{eff} \\
\delta_2 &=& ({k_F d \over 2\pi^2} - 1) \delta_1~. 
\eea
Here $n$  and $m$ are the number  of electrons  on the $`+/-'$ wires
when they are off resonance and the $`-/+'$ wire is tuned to resonance.
Since the resonance condition of one wire (say wire A)  depends on the
number  of  electrons of the other wire (wire B), 
unlike the case when the two wires 
are decoupled (${\cal U}_1={\cal U}_2$), wire B
clearly has to have a fixed  number of electrons - $i.e.$, it has
to be off-resonance. (At resonance, the wire is degenerate for 
$n$ and $n+1$ electrons and hence does not have a fixed number
of electrons. The electron number fluctuates.) 
So the derivations of the gate voltages above for the $`+/-'$
wires are valid only when the $-/+$ wire is far from resonance. 
Our analysis is unable to predict conductances for 
${\cal U}_1\ne {\cal U}_2$ when both
the wires are near resonance.

\section{Effective action of coupled dots}

In this section, we map the effective action studied in the
earlier section to the effective action of capacitively coupled
quantum dots to obtain the conductance pattern for coupled
dot systems.

We note that Eq.(\ref{ea}) is precisely the effective action
of coupled quantum dots with charging energies ${\cal U}_{eff}$
and an interaction energy $({\cal U}_1 -{\cal U}_2)/4$. 
In the absence of the interaction term, $i.e.$, when ${\cal U}_1 =
{\cal U}_2$,
by tuning $N_{0\nu}$ or equivalently by tuning $g_i$
the dot states with $N_\nu$ and $N_\nu+1$ can be made degenerate. 
This is the lifting of the Coulomb blockade (CB)  for each
individual dot. The gate voltages at which both the CB's
are lifted and the current through both dots is at a maximum
are the same points in Fig. 3 where both the wires go
through a resonance. Similarly, the gate voltages where one of 
the CB's is lifted is where one of the wires goes through a
resonance.

\begin{figure}[htb]
\begin{center}
\epsfig{figure=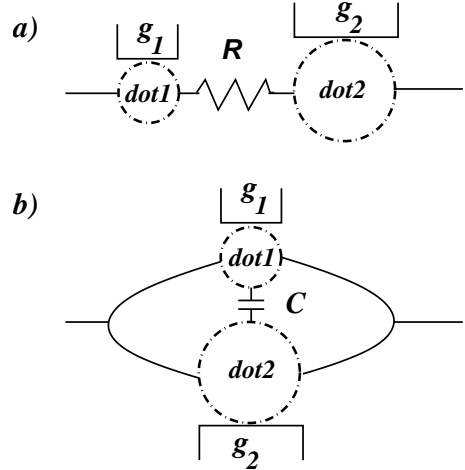,width=6.0cm}
\end{center}
\caption{Schematic diagram of  (a) tunnel coupled 
quantum dots  in series and (b) capacitively coupled  quantum dots in 
parallel.}
\end{figure}

Note that although the effective action looks similar to the
effective action for tunnel coupled quantum dots\cite{us},
there is an important difference. Unlike the tunnel-coupled case (Fig. 4(a)),
here, we have a two channel problem (Fig 4(b)).  Hence, there is non-zero
conductance even when only one of the Coulomb blockades
is lifted. For instance, when the two dots are weakly
capacitively coupled, ($U_1-U_2$ is small), we can trivially see that 
when the CB through dot 1 is lifted, 
$G_1 \ne 0$ and when the CB through dot 2 is lifted, 
$G_2 \ne 0$.
Thus, if we measure the total conductance through both the dots, 
the lines of semi-maxima
are when one of the CBs is lifted and the points of 
maxima are when both CBs are lifted. In contrast, for the tunnel
coupled dots, the maxima occur only when both Coulomb blockades
are lifted. 

\begin{figure}[htb]
\begin{center}
\epsfig{figure=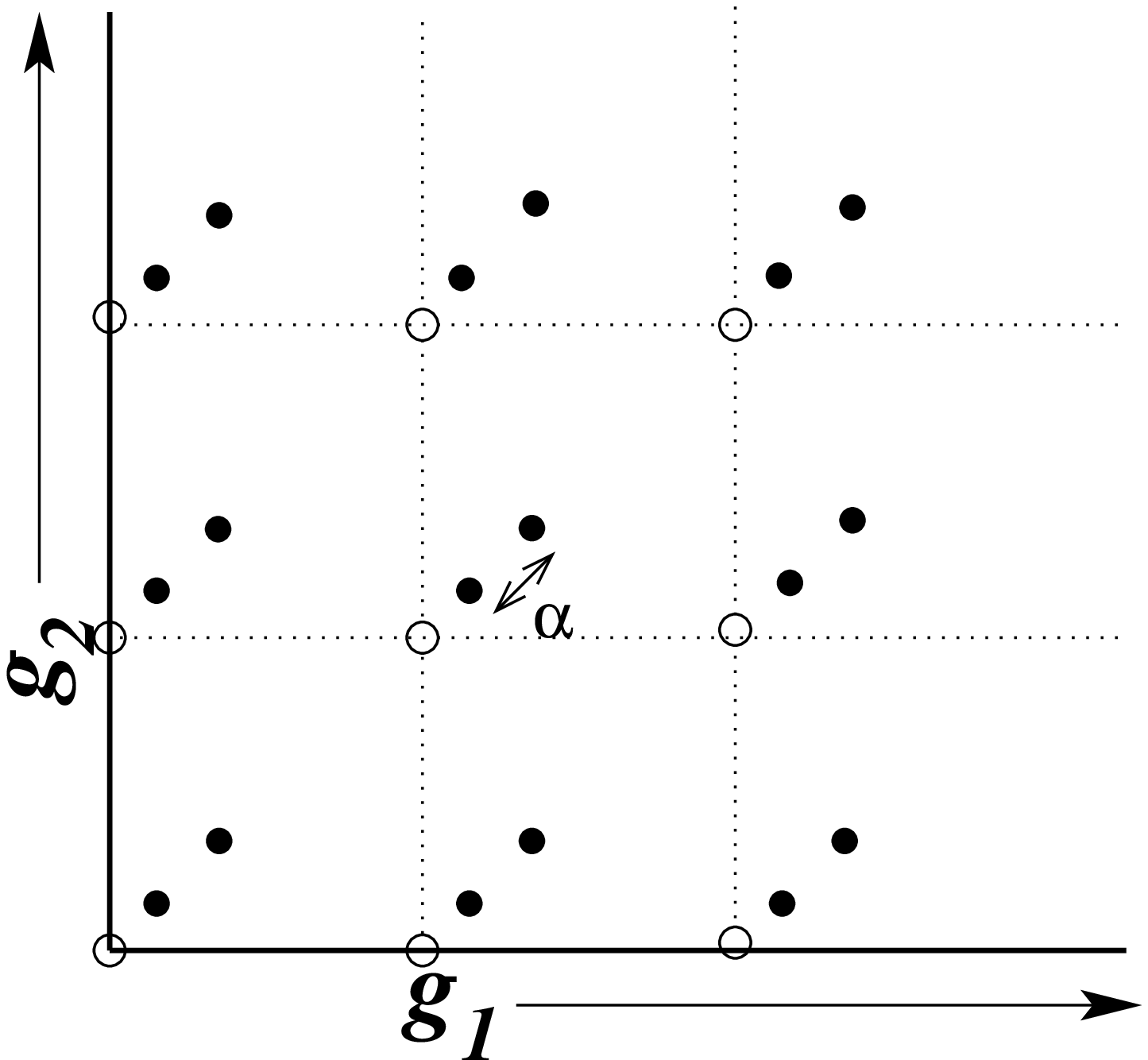,width=6.0cm}
\vspace{1.0cm}
\epsfig{figure=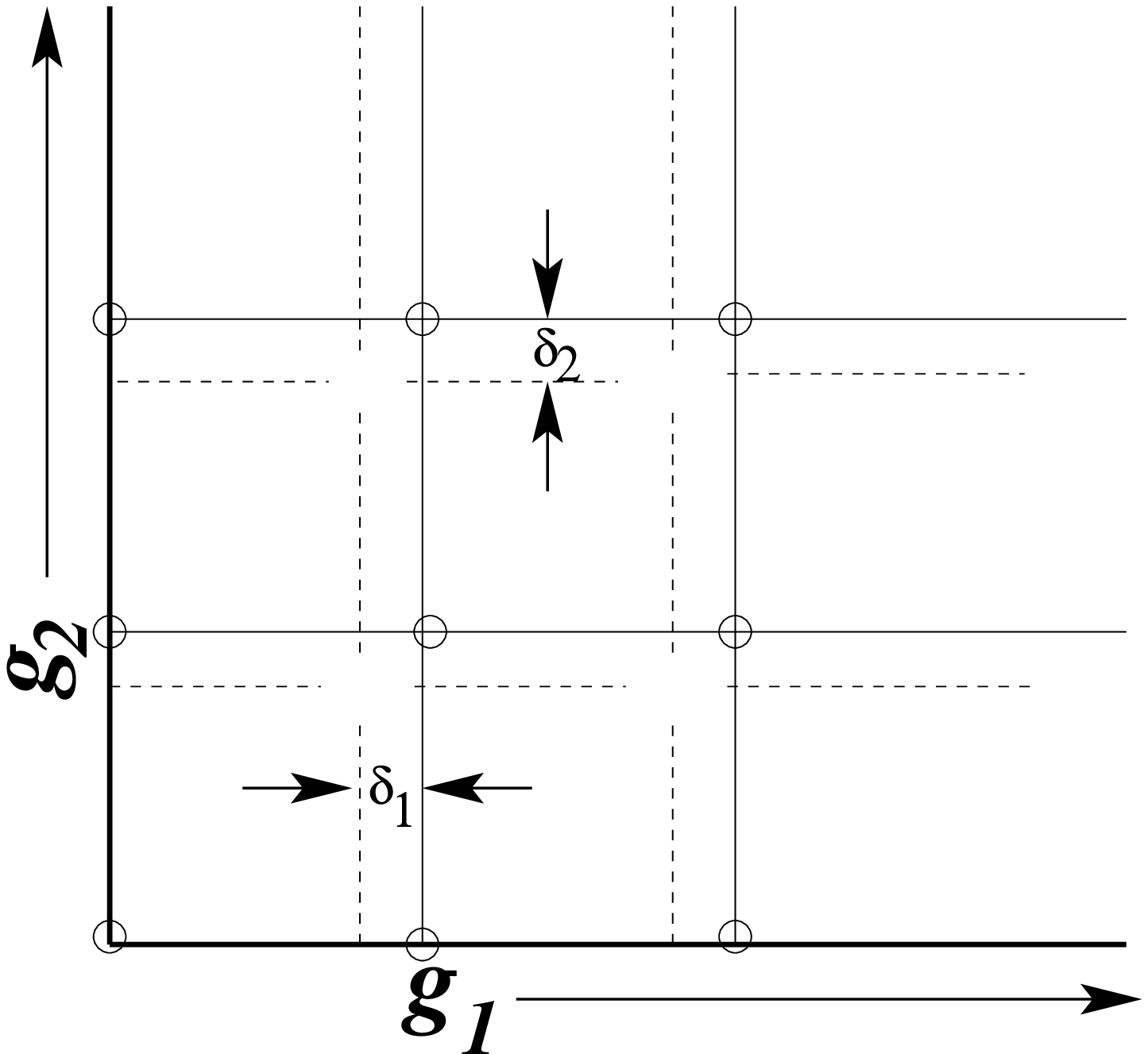,width=6.0cm}
\end{center}
\caption{Conductances in the plane of the gate voltages $g_1$ and $g_2$
for (a) tunnel coupled dots and (b) capacitively coupled dots. In the
absence of coupling between the dots, the resonance maxima  form
a square grid in both cases as shown by the open circles. 
In (a), the open circles  are the only points where
there is a maxima, because both dots need to be at resonance. 
In (b), the solid lines indicate semi-maxima (where one of the dots is on
resonance) and the open circles  denote maxima where both dots
are at resonance. When interdot coupling is introduced, in (a), each point 
of resonance splits into
two as shown by the solid circles. In (b), the lines of semi-maxima
shift as shown by the dotted lines, and there are no points where
both dots are at resonance.}
\end{figure}                   

When ${\cal U}_1 \ne {\cal U}_2$, we have a
term mixing $N_1$ and $N_2$. This  tells us that the CB through one
dot is affected by the charge on the other dot.
As in the case of wires, this 
means that the CBs through both 
dots cannot be simultaneously lifted . The lines  where one of
the CBs is lifted is shifted from the ${\cal U}_1={\cal U}_2$ case and
as for the wires, we are unable to predict conductances at the
crossing points. In contrast,  the effect of a weak interdot coupling 
in the tunnel coupled case is
to split the maxima\cite{exp,us}. These results are depicted in Fig. 7,
to show the contrast.

\section{Discussions and conclusions}

In this paper, we have studied conductance through a 
pair of carbon nanotubes, which are coupled by floating
gates at the beginning and end of the wires. This geometry
of carbon nanotubes enables us to study how resonant tunneling
conductance through one carbon nanotube is affected by that
of the other. We have obtained the conductance pattern
as a function of the two gate voltages controlling the
densities of the electrons in the two wires. 
In the plane of the two gate voltages, we find that 
(for identical carbon nano-tubes), 
the conductance is a semi-maximum (goes through
a single resonance) along the lines 
$g_2-g_1 =-(m+1/2) {\cal U}$ and $g_1+g_2 =
\big(2 k_F d/{\pi}) - (n+1/2)\big) {\cal U}$. 
At the points where the two lines 
cross, the
conductance is a maximum (goes through two resonances).
In the rest of the plane, the conductance  
is very low  (no resonance).
When the two wires are not identical, the lines of 
semi-maximum (single resonance)
shift to $g_1-g_2 =\big[(2 k_F d/{\pi})\big(({\cal U}_1 -
{\cal U}_2)/({\cal U}_1 + {\cal U}_2)\big) - (n+1/2) - 
(\Pi/4)\big(({\cal U}_1 -{\cal U}_2)/({\cal U}_1 + 
{\cal U}_2)\big)\big]  2{\cal U}_{\cite{kbf,eg,many} eff} $ 
and $g_1+g_2 =\big[(2 k_F d/{\pi}) - (m+1/2) - 
(\Pi/4)\big(({\cal U}_1 -{\cal U}_2)/({\cal U}_1 + {\cal 
U}_2)\big)\big]  2{\cal U}_{eff}$ 
and there is no resonance when the lines cross. 
We have also
mapped the problem to that of two quantum dots that
are capacitively coupled. The conductance through the
double-dot system shows the same patterns of maxima
when both the dots are on resonance, a semi-maxima
when one is on resonance and no conductance otherwise.

The above analysis has been a low temperature analysis, 
$T \ll  T_d = \hbar v_{eff}/ K_{eff} d \sim 1 K$ for typical
wires of length $d=5\mu m$. This is needed for  
coherent propagation through the wire leading
to resonance features. 
Hence, it is the length $d$ which
plays the role of a cutoff in the RG
flows. Although, it may not be experimentally feasible
to change the lengths of the wire, if it could
be done, then one would expect the deviations of
conductances from perfect resonance to scale as power laws
of the lengths, as usually happens in LLs.
The inclusion of leads also brings in non-trivial temperature
dependences even in the low temperature limit. These may
be experimentally easier to see.
Thus such experiments would probe Luttinger liquid physics.
More importantly, the geometry that we have studied also
allows for cross-conductances, whose temperature and
length dependences
also show the characteristic LL power laws. Here, however,
the very existence of a `cross' current is an interaction 
dependent effect and thus provides a qualitative probe
of LL physics.

Qualitative tests of the other features that we have studied
should also be experimentally feasible.  Conductances
through a double wire system or a capacitively coupled
double dot system, should show the features that are  seen in 
Figs. 3 and 7. There should be large differences in the conductance
in the three different cases where 1) 
both the gate voltages are tuned to resonance
(maxima of conductance) 2) when one of them is tuned
to resonance (semi-maxima of conductance) and 3)
when both are out of resonance ( very low conductance). 

Finally, this analysis can be easily extended to the case
where the two wires are allowed to `cross' at more than two points.
A very similar analysis
shows that the system can still be decoupled in terms of $`+'$
and $`-'$ wires, in terms of which the problem reduces to that of
Luttinger wires with multiple barriers\cite{us}.
For three crossings,  the equivalent dot geometry involves four dots, 
at the four corners of a square, with tunnel couplings along
the horizontal axis and capacitive couplings along the vertical
axis. These are also  interesting geometries to study\cite{wip}
in the context of quantum computers.

\vskip -0.6 true cm

\end{document}